# Vector Boson Fusion and Quartic Boson Couplings

**Dan Green**
**(dgreen@fnal.gov)**
**US CMS Dept.**
**Fermilab**

One of the goals of the upcoming experiments at the LHC is the exploration of electroweak symmetry breaking in all aspects. In particular, the self-couplings of the vector gauge bosons are completely specified in the Standard Model, as are the couplings of the Higgs boson to the vector bosons once the Higgs mass is known. The vector boson fusion mechanism is examined in order to measure the process W + W -> Z + Z which has a favorable signal to noise ratio and which has a major contribution from the quartic coupling WWZZ.

## Introduction

In the Standard Model (SM) there exist triple and quartic gauge boson couplings, fundamentally because the weak interactions are described by a non-Abelian gauge theory [1]. The only triple couplings allowed in the SM are WWZ and WW$\gamma$. There are no ZZZ, ZZ$\gamma$, Z$\gamma\gamma$, or $\gamma\gamma\gamma$ couplings in the SM, for example. At LEP2 WW and WW$\gamma$ final states have been observed [2]. These final state data are then used to infer the SM couplings WWZ, WW$\gamma$ and WW$\gamma\gamma$, WWZ$\gamma$ respectively. At the present level of accuracy no deviations from the SM predictions are found. These predictions should also be tested in the near future at the Fermilab Tevatron experiments, CDF and D0, at an increased level of accuracy using data containing WW, WZ, W$\gamma$, and ZZ in the final state.

The only allowed quartic couplings are WWWW, WWZZ, WWZ$\gamma$ and WW$\gamma\gamma$ in the SM. The strength and character of these couplings is completely specified in the SM. Clearly, these predictions should be tested as rigorously as possible. The most obvious approach is, by extension, to look at the production of three gauge bosons in the final state. In this note another approach is considered, using vector boson fusion. In that mechanism, the final state containing two vector bosons and two “tag” jets is a characteristic topology indicating virtual two-body vector boson scattering, which contains a Feynman diagram with quartic coupling. That study would be complementary to the study of the production of three gauge bosons in the final state where only one of the gauge bosons is virtual rather than the two in the initial state.

## VV Production

The SM triple couplings will presumably be well studied at the Tevatron. They will be explored using the “Drell-Yan” (D-Y) process of quark-antiquark annihilation. The Feynman diagrams generated by COMPHEP [3] for WW production are shown in Fig.1.a. Note that the WW final state in the D-Y process is due to diagrams containing the two triple couplings with an intermediate Z or $\gamma$. The mass of the WW pair falls smoothly and fairly slowly from threshold, as seen in Fig.1.b. which gives the COMPHEP prediction for the cross section for p-p scattering at 14 TeV, or the LHC.

In principle the quartic couplings can be studied and tested by exploring reactions with three vector bosons in the final state. For example, WWZ and WW$\gamma$ final states would probe the WWZZ, WWZ$\gamma$ and WW$\gamma\gamma$ couplings. Finally, WWW production, with an intermediate virtual W boson, or W*, would probe the WWWW coupling. All these couplings are completely specified in the SM and the predictions should be thoroughly tested. However, the rates are rather low and an accurate measurement may require a higher luminosity for the LHC than is expected during the initial operating period [4]. An alternative possibility is to use the vector boson fusion process to study two-body vector boson scattering.

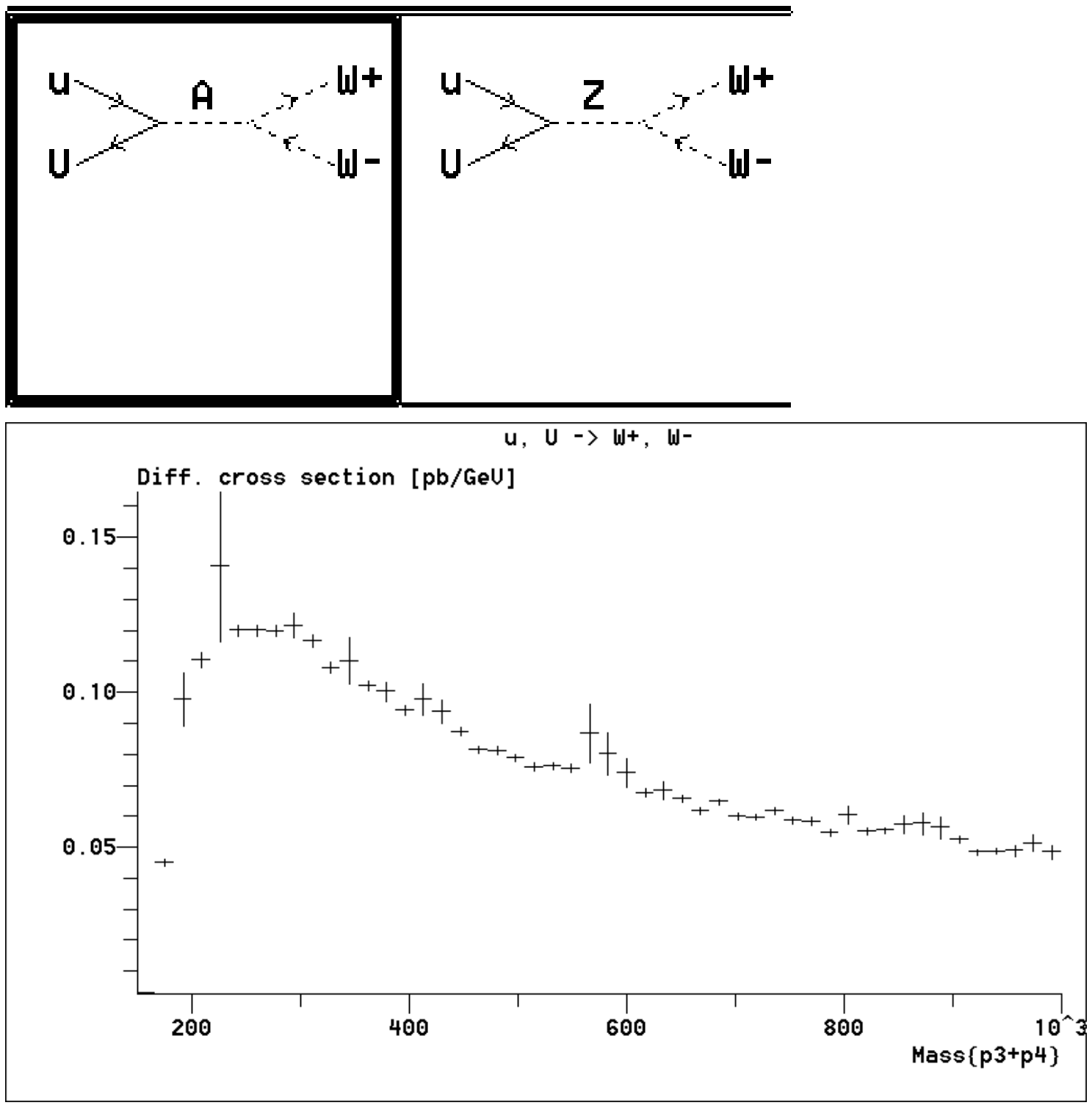

a)

b)

Figure 1: a) Feynman diagrams for quark-antiquark annihilation into a virtual photon or Z boson, which decays into W pairs. b) Mass spectrum for the Drell-Yan process shown in a) at the LHC.

Consider now the production of Z pairs in the SM. Because of the limited triple couplings allowed in the SM the ZZ final state, in lowest order, has only one allowed Feynman diagram, as shown in Fig.2.a. It represents not a Drell-Yan like process but rather double radiation of a Z by a quark in the absence of a D-Y mechanism. The ZZ mass spectrum falls very rapidly, as seen in Fig.2.b, because the Z are radiatively produced. The total cross section for this specific process is ~ 2 pb in 14 TeV p-p collisions (LHC). Considering that in p-p collisions we have four processes of roughly equal strength, ($u+\overline{u},\overline{u}+u,d+\overline{d},\overline{d}+d$), we estimate the full proton-proton cross section to be ~ 8 pb which agrees with other [1] estimates.

The rapid fall of the ZZ mass spectrum and the relatively small cross section at p-p colliders means that searches for the Higgs boson decaying into this final state enjoy a relatively good signal to background ratio. Indeed, these final states, with the Z decaying into leptons, are thought to be the “golden modes” for Higgs searches at the LHC [5].

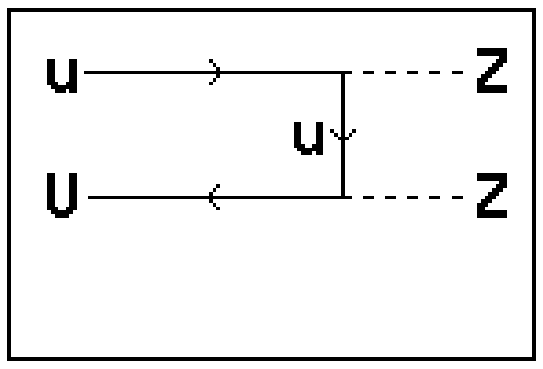

a)

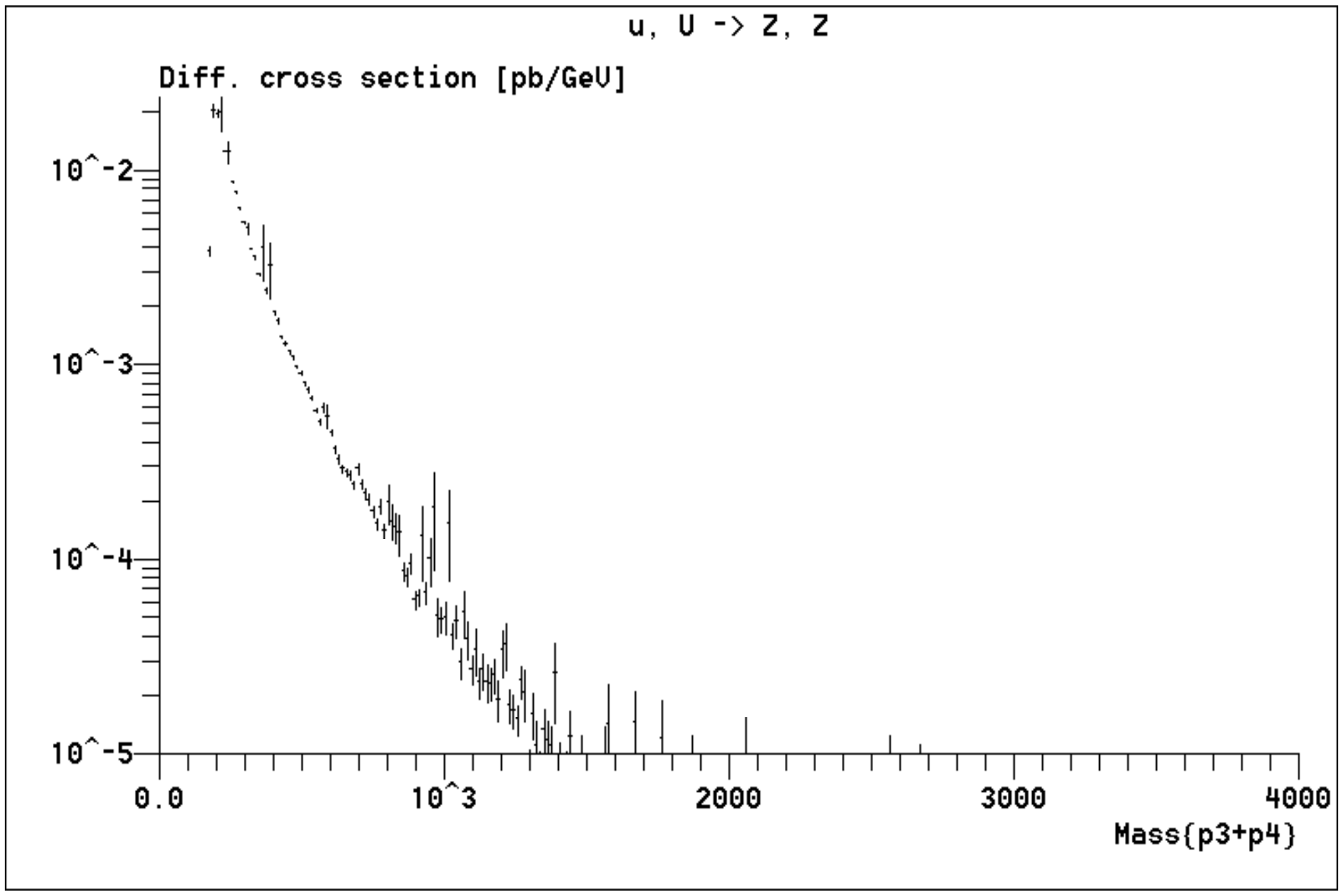

b)

Figure 2: a) Feynman diagram for ZZ production in p-p collisions in lowest order.
b) Mass spectrum for the ZZ pair at the LHC.

## VV Fusion in Higgs Production

In the recent past, the vector boson fusion process has been studied and found to improve the signal to background ratio in Higgs searches, e.g. in $\gamma + \gamma$ decays, sufficiently so that several new final states at low Higgs mass, such as $\tau^+\tau^-$ and WW* have become accessible [6,7]. For the Higgs, the vector boson fusion process allows one to explore the initial state WWH (and ZZH) coupling. Indeed, in the WW* final state, the full vector fusion reaction depends solely on the WWH coupling, which allows for a clean and direct measurement of the fundamental WWH coupling strength. In this manner, Higgs coupling to vector bosons can be isolated in the vector boson fusion process.

The Feynman diagrams for the production of a Higgs by vector boson fusion are shown in Fig. 3.a. Note that the topology of interest has two quark jets at large rapidities in the final state with a centrally produced Higgs. In addition, this process can be "calibrated" by using WW production of Z, assuming that the WWZ coupling is measured to be as predicted in the SM. Looking at Fig. 3.b, it is easily seen that Z production in vector boson fusion has exactly the same topology as Higgs production. In fact, this observation is very similar to that already made for associated production of Higgs with top pairs with subsequent decay of the Higgs into b quark pairs [8]. In that case too, Z production has exactly the same topology and a similar coupling strength. Since the Z is easily and cleanly reconstructed using lepton pairs, both ttH and jjH can be "calibrated" using the topologically and kinematically similar ttZ and jjZ final states. The notation of jj indicates the existence of two "tag" jets in the final state, which imply, or tag, the virtual emission of two vector bosons in the virtual initial state. The Z production process therefore provides a "standard candle" for vector boson fusion.

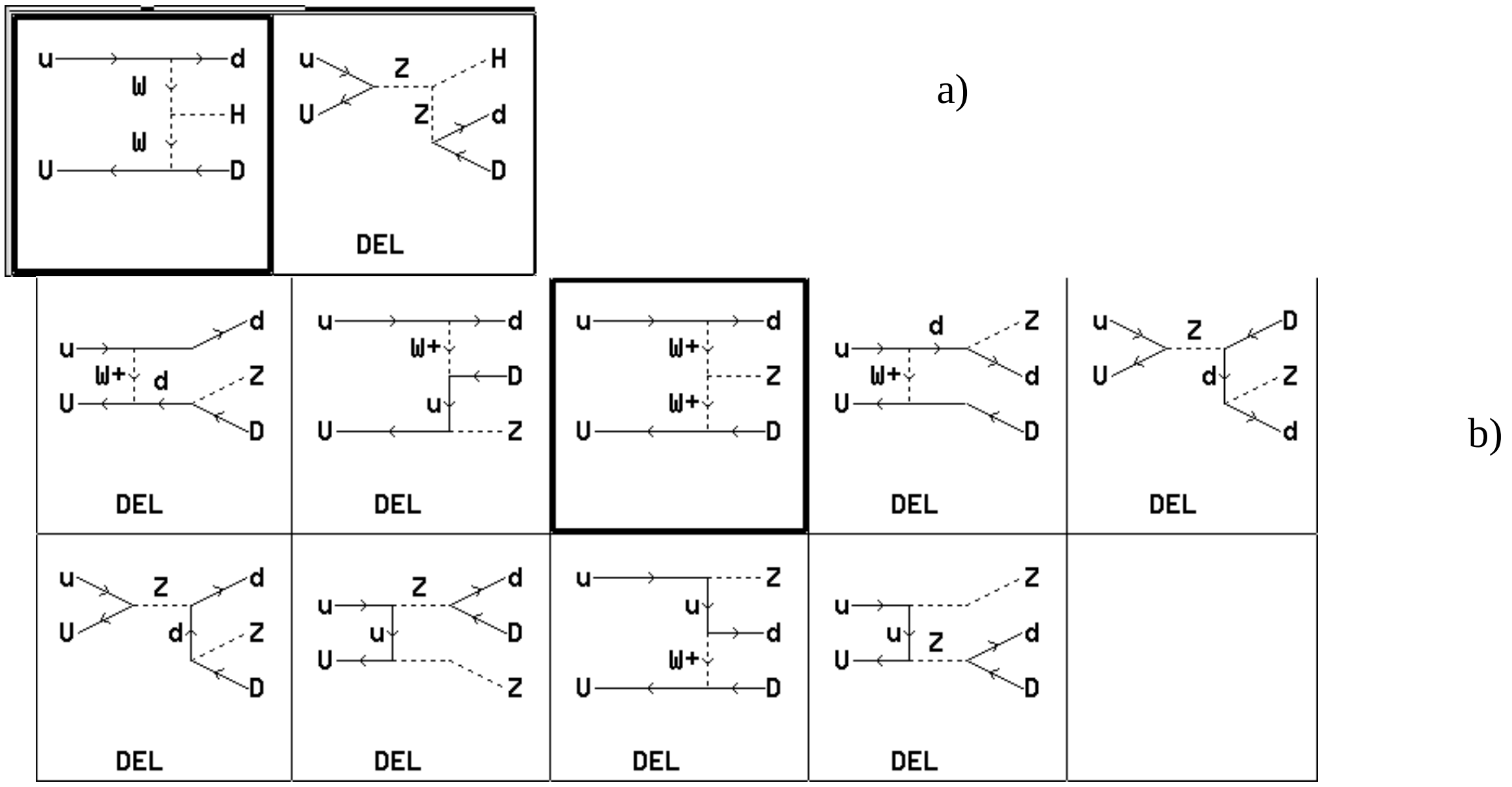

Figure 3: a) Feynman diagrams for the process $u + \bar{u} \rightarrow d + \bar{d} + H$ . b) Feynman diagrams for the related process $u + \bar{u} \rightarrow d + \bar{d} + Z$ .

The kinematics for vector boson fusion production of Z or Higgs bosons decaying into WW* and thence to dileptons plus neutrinos is essentially the same. There are two "tag" jets + central dileptons. The tag jets have transverse momentum ~ 40 GeV and rapidity <y> ~ 3. The Z or H is produced centrally. The transverse momentum of the Z or H is quite large, peaking at ~ 100 GeV, with a mean, $\langle P_T \rangle$ ~ 200 GeV. The large values of boson transverse momentum are characteristic of the vector boson fusion process. Therefore, trigger efficiencies will be high. The VV fusion cross section for Z (150 GeV H, as illustrated in Fig.12) production is ~ 26 (5) pb. Using muon and electron decays for the Z (W and W*), gives 1.7 pb for the dilepton final state cross section times branching ratio or 17,000 (1600) events in an initial run of 10 $fb^{-1}$ assuming perfect triggering and reconstruction efficiency. Therefore there should exist a good calibration sample of Z to use during early data taking at the LHC.

## Quartic Coupling in ZZ Production

Based on the promising Higgs production studies using the mechanism of vector boson fusion, it seems natural to ask about the role of quartic couplings in gauge boson pair production by way of vector boson fusion. Clearly, the two tag jet topology indicates a VV initial state so that a VVjj final state should indicate, with some background, a VV two body fundamental scattering. The ZZjj final state is chosen because, as discussed earlier, lowest order ZZ production, lacking a D-Y process, has a small cross section, limited to low ZZ masses.

Since the Higgs is meant to unitarize the VV scattering, studying this process to check for the predicted SM behavior is of some interest. Basically, there are three contributions to W + W -> Z + Z scattering. There is "s channel" virtual H production, followed by H -> Z + Z decay. There are also two diagrams for virtual W + W -> Z + Z scattering, one due to "t channel" W exchange and one due to direct WWZZ quartic coupling. Note that all three diagrams separately rise with increasing C.M. energy, indicating non-renormalizable behavior. It is only due to the cancellations among these three diagrams in the SM that a constant total cross section for W + W -> Z + Z is finally obtained. Clearly, checking for the validity of these postulated cancellations is of great importance.

The Feynman diagrams for the process $u+\overline{u} \rightarrow d+\overline{d}+Z+Z$ are shown in Fig.4. There are three diagrams which have the distinct topology with two tag jets. In those three diagrams there is a dependence on HVV, WWZ and WWZZ coupling. The HVV and triple couplings are assumed to be as predicted in the SM. These three diagrams are precisely those which occur in the fundamental process W + W -> Z + Z with the W being real external particles in that case.

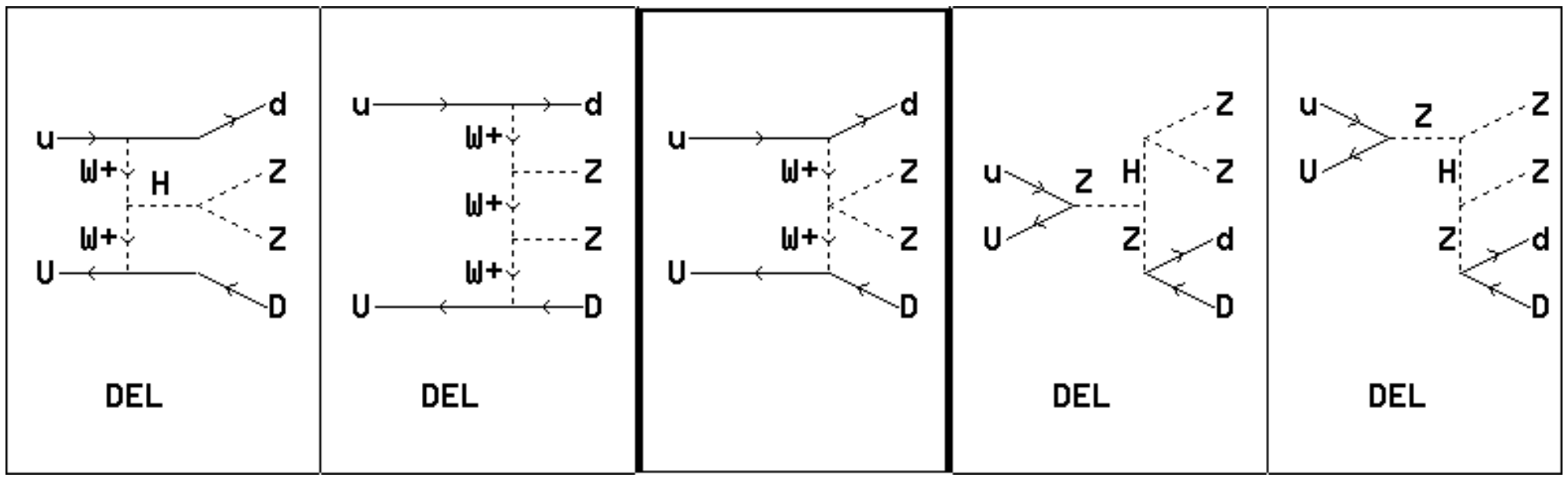

Figure 4: Feynman diagrams for the process $u+\bar{u} \rightarrow d+\bar{d}+Z+Z$ with some restrictions on allowed intermediate particles.

There is some discrimination among these three diagrams based on the "decay" angular distributions of the Z pairs. For the Higgs, a scalar will decay isotropically. In contrast, the quartic diagram, selected in isolation, gives a decay angular distribution going roughly as $\sim 1+\delta \sin^2\theta$, where the angle is that of the Z in the ZZ C.M. system, as seen in Fig.5.

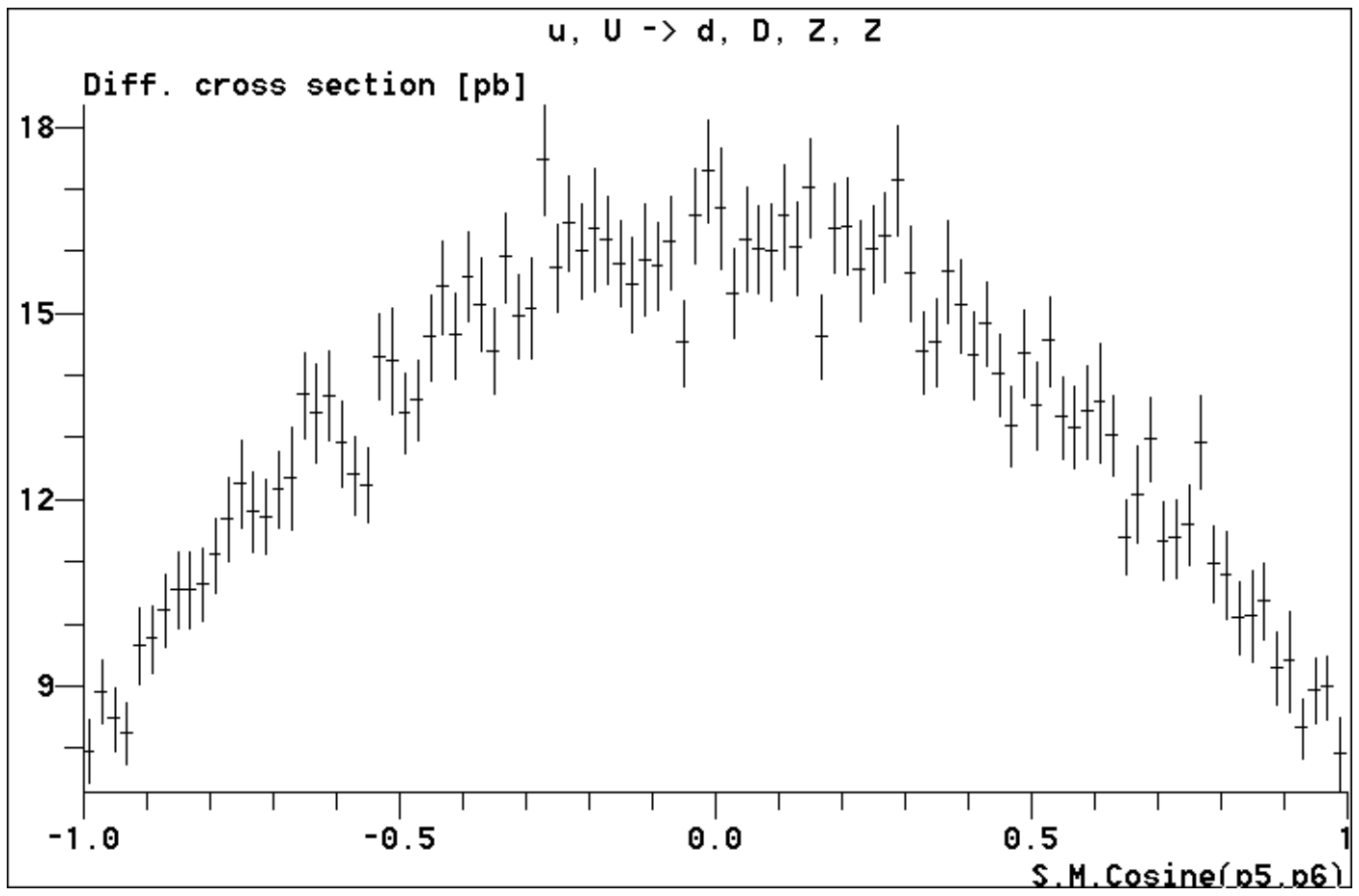

Figure 5: Distribution of the Z decay angle in the ZZ center of mass for the ZZ "quartic" diagram alone in p-p collisions at 14 TeV (LHC).

In the fundamental process, W + W -> Z + Z there is a strong cancellation between the isotropic Higgs decay and the $1-\varepsilon\cos^2\phi$ distribution for the W exchange diagram and the quartic diagram, which peaks at $\cos\phi = 0$, where the angle $\phi$ is between the incident W and the outgoing Z in the WW C.M. system. The complete reaction, summing all three diagrams, has an angular distribution which is very peaked in the forward/backward direction for W + W -> Z + Z.

Some of the kinematic characteristics of the tag jets and the ZZ pair are shown in Fig. 6 for the case when the three “tag jet” topological diagrams (three leftmost diagrams in Fig. 4) are active. The plots in Fig.6 refer to p-p collisions at 14 TeV (LHC). First, note that the high mass of the ZZ pair (Fig. 7) forces the rapidity of the tag jet to be not <y> ~ 3 as is the case in H production, but to have a more central rapidity. The Z in the final state are also, by kinematic constraint, forced toward rather central values of rapidity, y ~ 0. The “tag” jets need not now have a large separation in rapidity, as shown in Fig.6.c. Nevertheless, there is a residual tendency for these two jets to be emitted in opposite directions.

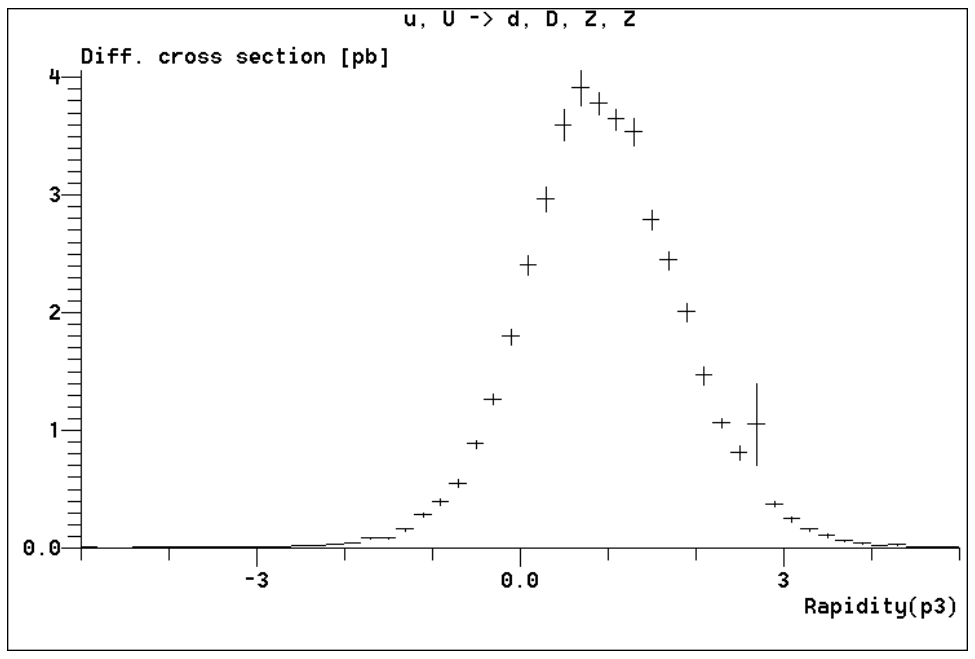

a)

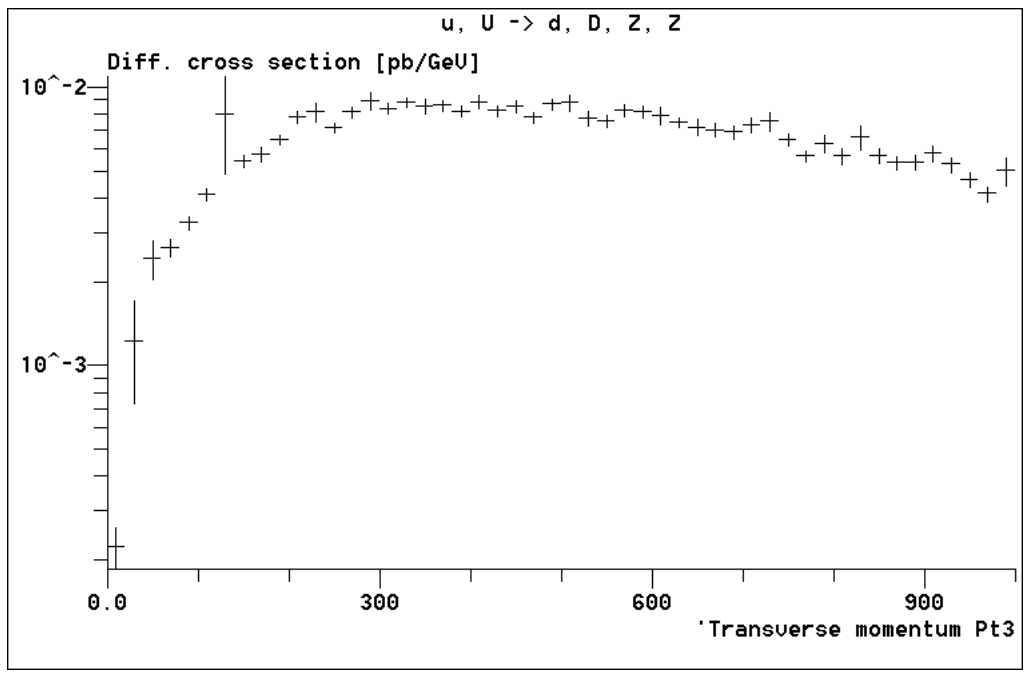

d)

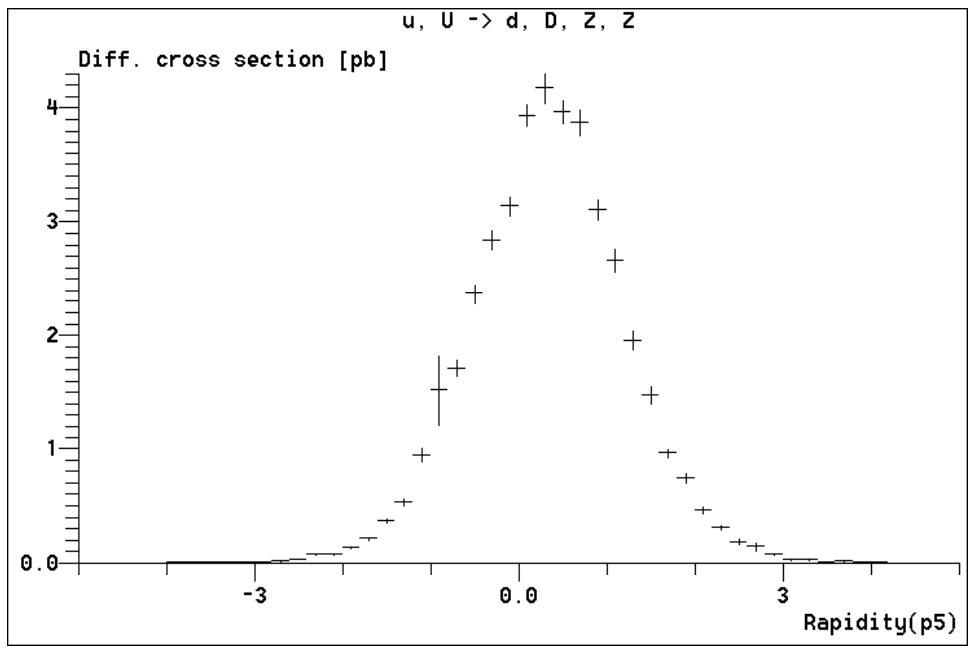

b)

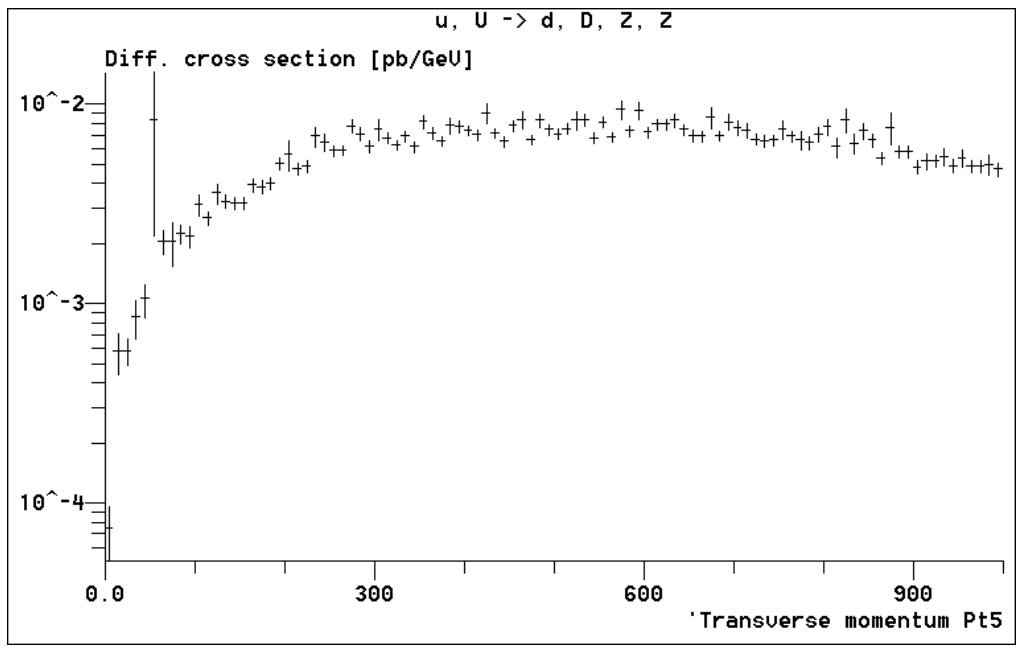

e)

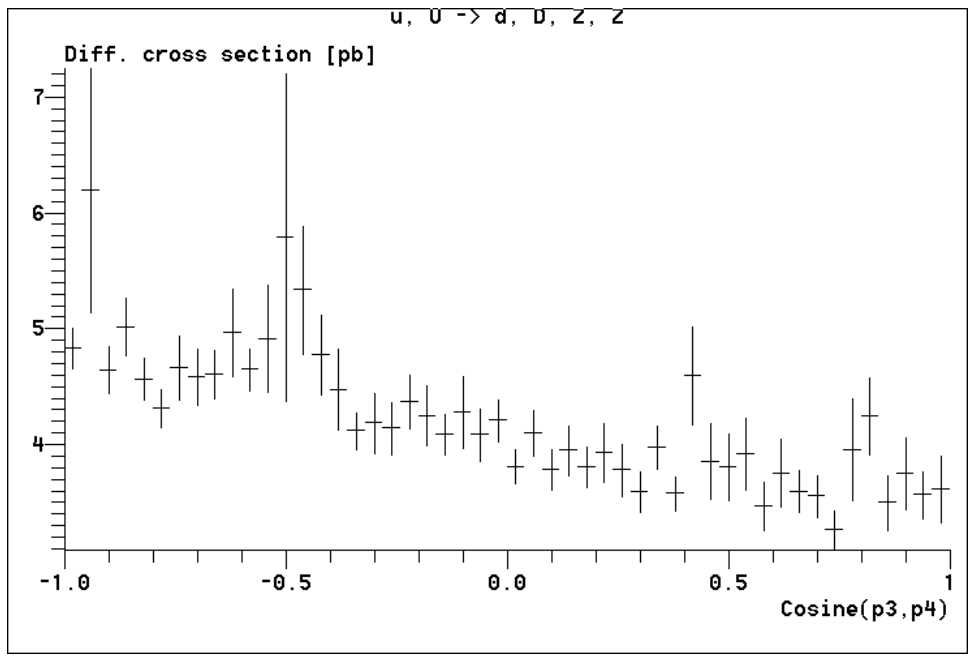

c)

Figure 6: Kinematic characteristics of ZZ production with associated tag jets in 14 TeV p-p interactions.

a) Rapidity of a tag jet
b) Rapidity of a Z boson
c) Cosine of the opening angle between the tag jets in the p-p C.M. frame
d) Transverse momentum of a tag jet
**e)** Transverse momentum of a Z boson

The mass spectrum for the p-p process at the LHC is shown in Fig.7. For this plot the Higgs mass is taken to be 185 GeV, so that the first bin at 200 GeV reflects the tail of the resonant Breit-Wigner Higgs mass distribution. Note that, even with the gauge cancellations required by the SM the spectrum extends up to ~ 2 TeV in mass. This spectrum should be compared to that for the lowest order process shown in Fig.2. Note also that the cross section is not small. The cross section for the single process in p-p interactions at the LHC is ~ 8 pb. For the p-p reaction (adding $u+\overline{u}, d+\overline{d}, \overline{u}+u, \overline{d}+d$) it should be ~ four times larger or ~ 32 pb.

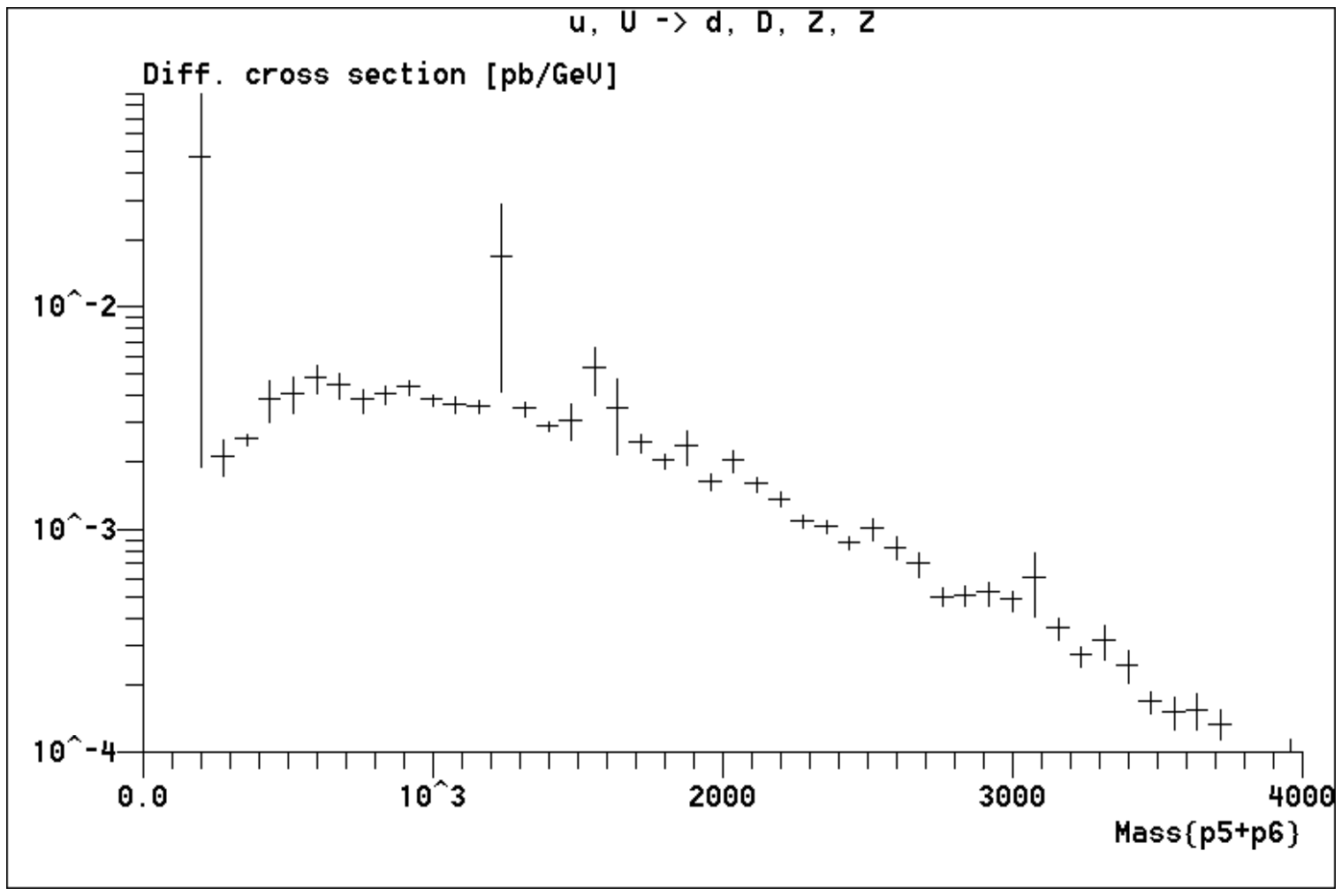

Figure 7: Mass of the ZZ pair in the vector boson fusion reaction at the LHC. The three leftmost diagrams shown in Fig. 4 are summed and squared in this calculation.

## W+W -> Z + Z and Quartic Coupling

The size and mass distribution of the ZZjj reaction is perhaps surprising. In particular, it exceeds the ZZ production cross section (Fig.2.b). Therefore, an analytic cross check was initiated. The starting point is the fundamental vector boson scattering. The COMPHEP result for the fundamental process is shown in Fig.8. The SM prediction is that the three diagrams all blow up with C.M. energy, but that the full process itself yields an energy independent cross section at high energies (well above the Higgs resonance) of ~ 328 pb. The structure at low ZZ mass is the tail of the Higgs resonance, assumed here to have a mass of 185 GeV. The magnitude of the cross section does not depend on the Higgs mass. It is dimensionally of a magnitude,

$\sigma(WW->ZZ) \sim \alpha_W^2 / M_W^2 \sim 67\ pb$.

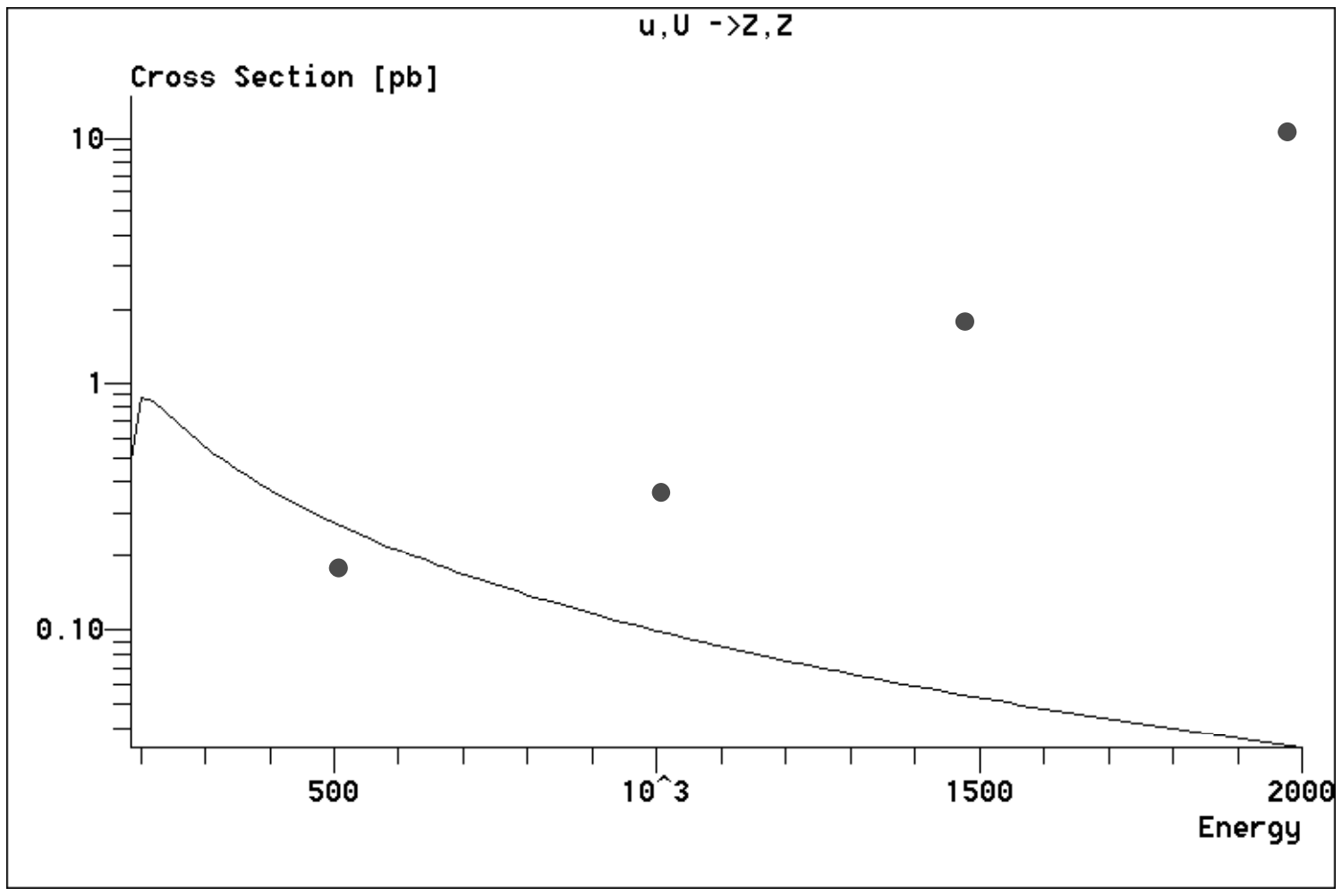

Figure 8: Cross section as a function of the C.M. energy for the fundamental process W+W->Z+Z. The cross section at high energies approaches a constant independent of the Higgs mass.

Moving from the fundamental process to the quark-antiquark initiated process, the production of Z + Z and $Z + Z + d + \bar{d}$ is examined. In Fig.9 is shown the cross section for the two processes. Clearly, the lowest order Z + Z process dominates at low ZZ pair mass, but the higher order process, $Z + Z + d + \bar{d}$, dominates at high ZZ pair mass. This dominance is due to the fact that there is no Drell-Yan process accessible to produce a Z+Z final state while there is a large cross section due to W + W -> Z + Z scattering which is available to the vector boson fusion process.

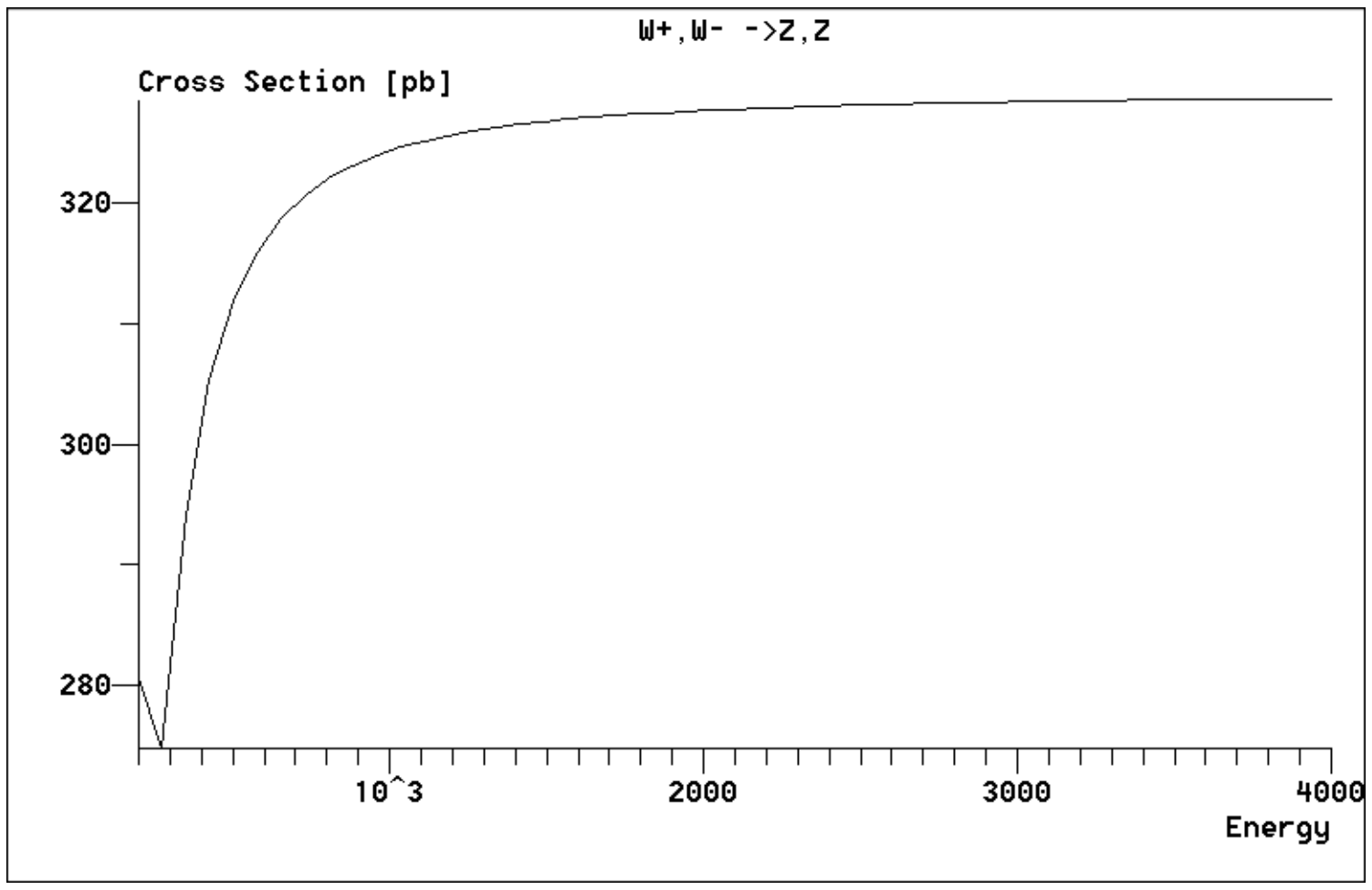

Figure 9: Cross section for the production of ZZ pairs in $u + \bar{u}$ collisions as a function of the C.M. energy in the initial state. The solid line is the cross section for the process shown in Fig. 2.a. The dots indicate the cross section for the processes indicated in Fig. 4.

The rise in the cross section for the higher order process is evident in Fig.9. This behavior can readily be understood in the context of the "effective W approximation" to the fusion process. The quark is, in this approximation, considered to be a source of virtual W bosons with a theoretical calculable probability. The transverse W dominate over longitudinal W in the source functions [5]. The effective luminosity, L, for the formation of the initial W+W state can be expressed analytically [5] as quoted in Eq.1.

$$(dL/d\tau)_{q\bar{q}/W_TW_T} = (\alpha_W/8\pi)^2(1/\tau)[\ln(\hat{s}/M_W^2)]^2[(2+\tau)^2\ln(1/\tau)-2(1-\tau)(3+\tau)] \quad 1)$$

The luminosity in quark-antiquark collisions depends on the electroweak coupling constant, $\alpha_W$, the C.M energy of the quark-antiquark, $\hat{s}$, the W mass, and the mass of the WW system = M. The parameter $\tau = M^2/\hat{s}$ largely defines the luminosity due to the strong $1/\tau$ dependence of L.

In the idealized case where the cross section for the virtual W+W process is constant with energy (e.g. Fig. 8, noting the suppressed zero for the vertical scale), the integral over the effective luminosity can be done analytically, and the quark-antiquark cross section can be given in closed form. The result, in the effective W approximation, is shown in Fig.10. The rise of the cross section with quark-antiquark C.M. energy is the result of the logarithmic dependence of the transverse W distribution function on quark-antiquark C.M. energy, as is evident in Eq.1. The magnitude of the result shown in Fig.10 and Fig.9 agree within a factor of two at a C.M. energy of 1 TeV, in addition. Therefore, the energy dependence of the exact COMPHEP calculation, Fig.9, can be understood in the approximation that the quarks are considered to be sources of virtual transversely polarized W bosons. The longitudinal contributions are expected [5] to be much smaller and are not considered here. Basically the longitudinal luminosity formula, analogous to Eq.1, lacks the crucial factor $[\ln(\hat{s}/M_W^2)]^2$.

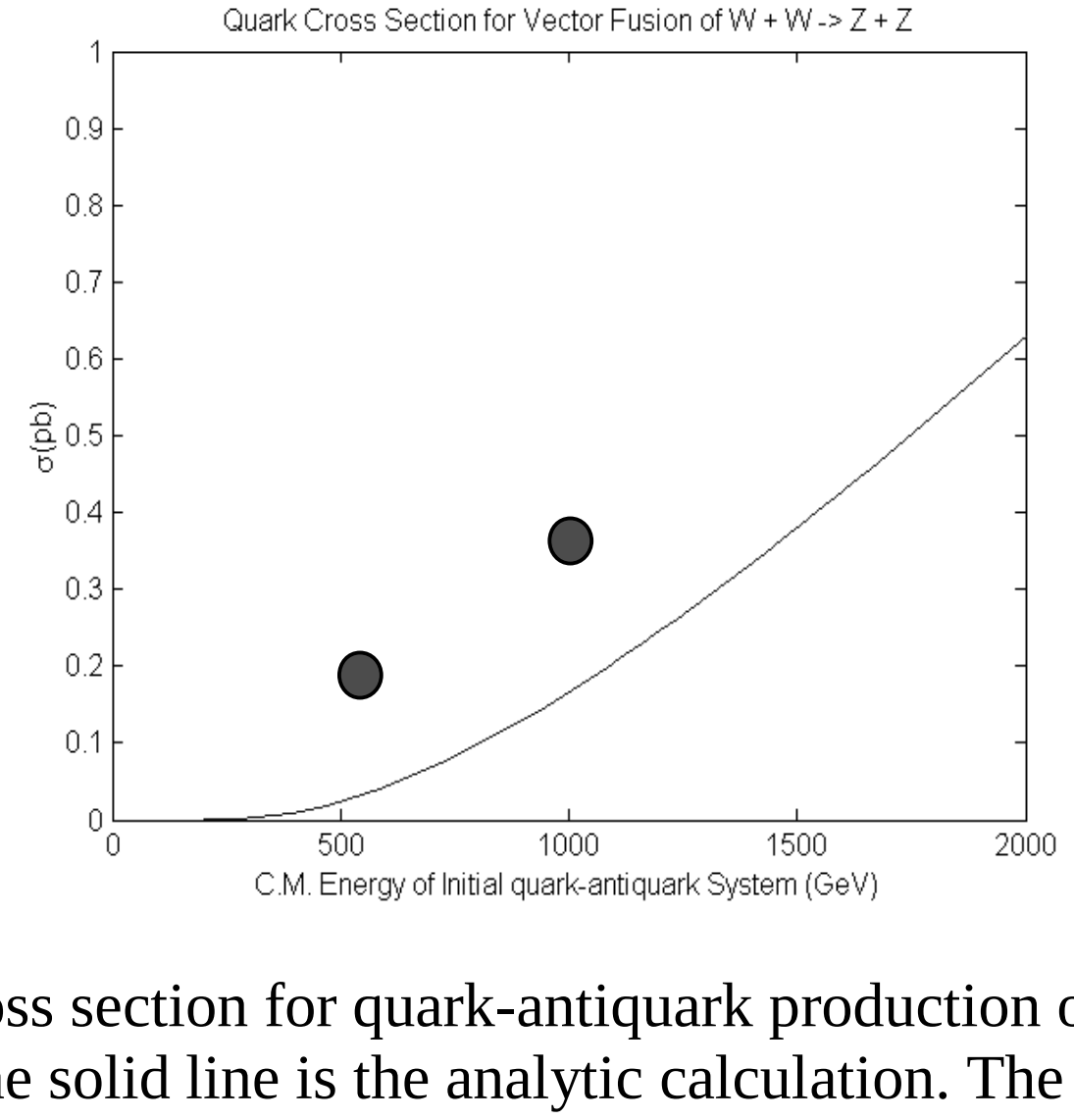

Figure 10: The cross section for quark-antiquark production of Z+Z in the effective W approximation. The solid line is the analytic calculation. The dots are the COMPHEP results taken from Fig.9. Note the rapid rise of cross section with C.M. energy.

The SM cancellations due to the interference of the different Feynman diagrams among the basic scattering diagrams can be examined in a very crude first pass by looking at the case where only the quartic diagram is allowed to contribute to the vector boson fusion process. At a fundamental level, this leads to a cross section which increases without limit at high mass. Indeed, the Higgs was invented in part precisely to solve this problem. At the level where we have experimental access to the process, p-p scattering at the LHC with tag jets indicating vector boson fusion, the cross section rises with ZZ mass (Fig.9) until it is pulled down by the falling parton distribution functions for the initial quark and antiquark. Nevertheless, although the fundamental divergence is no longer evident, in the case of a purely quartic Feynman diagram the cross section is much larger and the ZZ mass spectrum extends to higher ZZ masses than in the case of the SM with destructive interferences between the three diagrams. This result is shown in Fig.11.

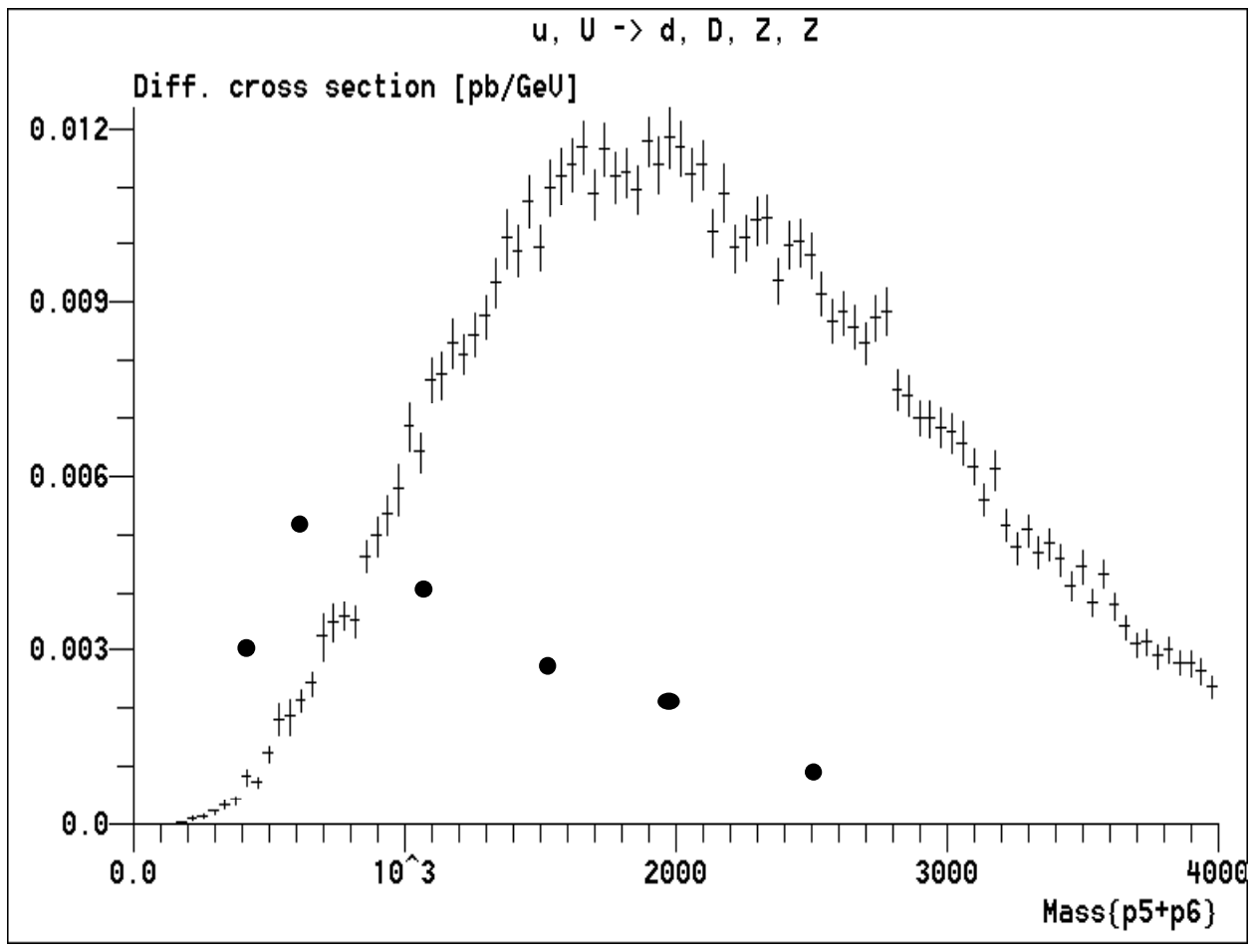

Figure 11: Spectrum of ZZ mass in p-p collisions at 14 TeV (LHC) for the case of vector boson fusion where only the quartic diagram is allowed. The dots are the result when the full SM prediction is evaluated (Fig. 7).

## Summary

The lowest order production of ZZ pairs has a cross section at the LHC of ~ 8 pb. In contrast, the production of ZZ pairs by vector boson fusion has a larger cross section at the LHC of ~ 32 pb. The higher order cross section also extends to much higher ZZ pair masses. The size of the cross section, the shape of the mass spectrum and the ZZ angular distribution each contain useful information. As such, the vector boson fusion process allows for an exploration of the predicted SM diagrammatic cancellations in the fundamental process, W + W -> Z + Z.

Previous studies [9, 10, 11] have concentrated on isolating $W_L + W_L$ scattering in vector boson fusion, W + W -> W + W, and exploring limits on non Standard Model interaction terms. A brief look at the rates for other processes with a dependence on a quartic coupling was made to explore the signal and lower order background rate issues.

For $\gamma + \gamma$ production in vector boson fusion the signal rate is rather low. For the case of $\gamma$ + Z production in vector boson fusion the cross section at the LHC is ~ 200 fb and exceeds the lowest order production rate for di-boson masses above ~ 1 TeV. In the case of W + W production in vector boson fusion the cross section is quite large. For the WWjj signal at the LHC the cross section is ~ 26 pb. The signal rate exceeds that for the lowest order Drell-Yan process (Fig.1) for WW masses > 0.7 TeV. As can be seen from Fig.7, the situation for WWjj is thus comparable to that for the ZZjj final state. Clearly, the data rates are high in the leptonic decay modes for both W. This process can be studied in more detail at a later time. In this work the ZZjj final state was looked at in some detail because it afforded a clean reconstruction, if leptonic decays are studied with two resonant Z masses. In the WWjj case, no invariant mass can be reconstructed.

The Higgs boson is largely produced at the LHC by the mechanism of gluon-gluon fusion [11]. Nevertheless, the higher order process of vector boson fusion is important at the LHC and indeed, becomes increasingly important as the mass of the Higgs particle increases, as is seen in Fig.12. At a 1 TeV Higgs mass, the production cross section for the two processes is almost the same.

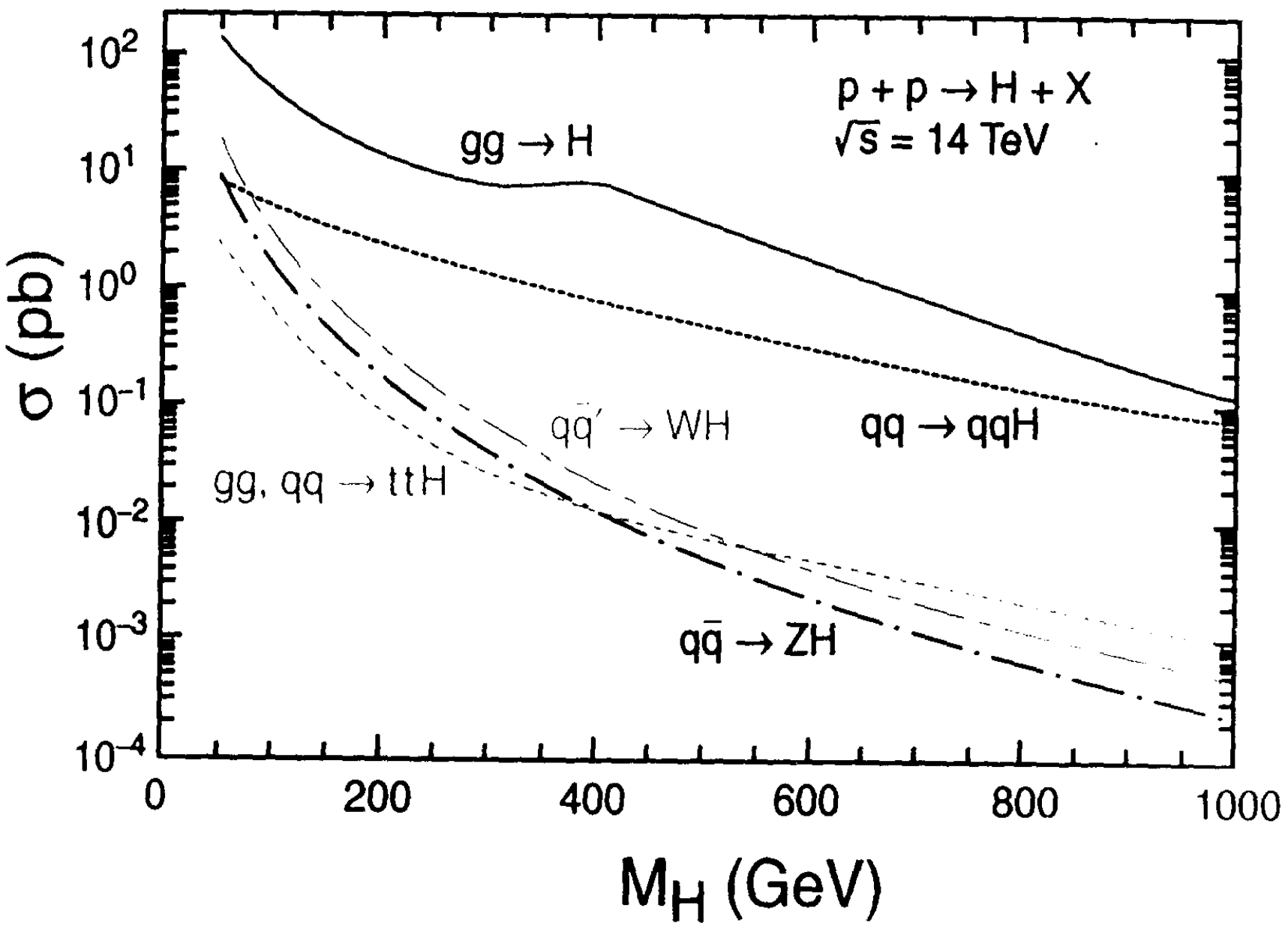

Figure 12: Cross section for production of the Higgs boson at the LHC. Several production mechanisms are indicated, where the two most important are gluon-gluon fusion and vector boson fusion.

The situation for ZZ pair production is summarized in Fig.13. The plot is for p-p collisions at the LHC. The histogram indicates the ZZ mass spectrum due to vector boson fusion (as in Fig.7). The dots represent the lowest order production of ZZ pairs (Fig. 2). The open circles are a crude estimate of the Higgs signal in the ZZjj final state. What is plotted is the cross section for Hjj from Fig. 12, divided by the natural Higgs width and then divided by four (to roughly go from the full p-p interactions to the $u+\bar{u}$ fraction of p-p scattering at 14 TeV). Another factor of three ( H -> ZZ branching fraction) has been suppressed in the interest of visibility. Thus the open circles crudely represent the size of the resonant Higgs "mass bump", scaled up by a factor of three.

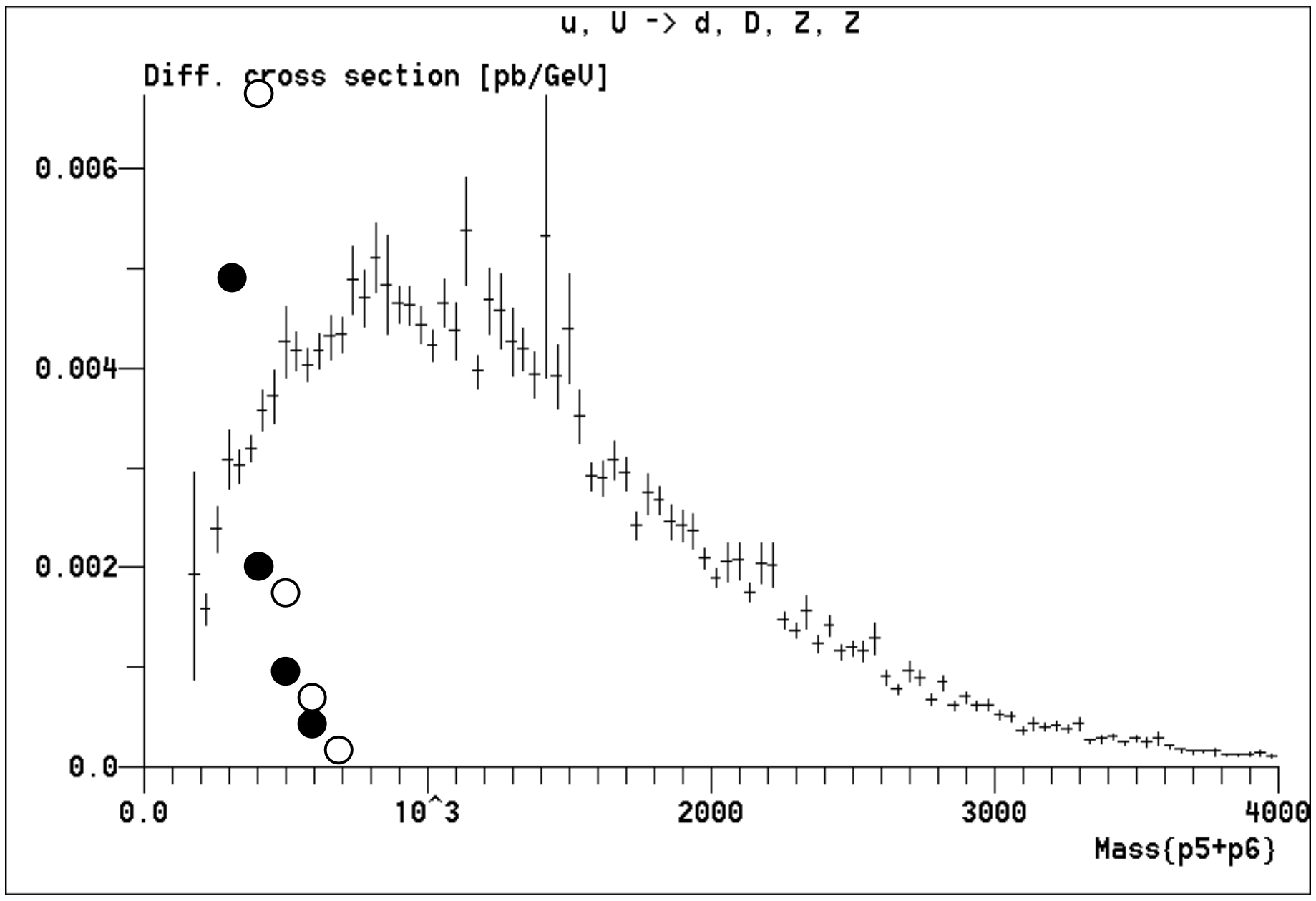

Figure 13: Cross section at the LHC for vector boson fusion production of ZZ pairs. The closed circles indicate the cross section for lowest order ZZ pair production, while the open circles represent the Higgs "mass bump" for H -> ZZ decays (times ~ three).

Clearly, the vector boson fusion source of ZZ pairs, ZZjj, is the dominant one for ZZ masses above ~ 350 GeV. They will constitute an irreducible source of background in the ZZ final state. That background will make high mass Higgs searches somewhat more difficult than has been previously thought. Perhaps a jet veto on the "tag" jets will be able to reduce this source to a sufficiently small value. Note however, that heavy Higgs production also has associated "tag" jets in the case of vector fusion Higgs production and that production is not negligible (Fig.12).

## References

1. V. Barger and R. Phillips, Collider Physics, Addison-Wesley Publishing Co. (1987)

2. LEPEWWG/TGC/2002-01 (2002)

3. A. Pukhov et al., User's Manual, COMPHEP V33, Preprint INP-MSU 98-41/542

4. F. Gianotti et al., CERN-TH/2002-078 (2002)

5. J. Gunion, H. Haber, G. Kane, S. Dawson, The Higgs Hunter's Guide, Addison-Wesley Publishing Co., (1990)

6. R. Rainwater, M. Spira, D. Zeppenfeld, hep-ph/0203187

7. S. Abdullin et al., CMS Note, 2003

8. D. Green, K, Maeshima. R. Vidal, W. Wu, CMS Note 2001/039

9. J. Bagger et al., Nucl. Phys. B399, 364 (1993)

10. J. Bagger et al., Phys. Rev. D, 49, 3, 1246 (1994)

11. J. Bagger et al., Phys. Rev. D, 52, 7, 3878 (1995)

12. Z. Kunst et al., Z. Phys. C 74, 479 (1997)